\documentclass[twocolumn,amsmath,amssymb,aps,superscriptaddress]{revtex4-2}
\usepackage{amsmath,amssymb, stmaryrd}
\usepackage{bm}
\usepackage[dvipdfmx]{graphicx}
\usepackage{ascmac}
\usepackage{float}
\usepackage{url}
\usepackage{comment}
\usepackage{tabularx,theorem}
\usepackage{indentfirst}
\usepackage{algorithmic}
\usepackage{algorithm}
\usepackage{enumerate}
\usepackage{color}
\usepackage{physics}
\usepackage{mathtools}

\usepackage[colorlinks=true,citecolor=blue,linkcolor=magenta]{hyperref}

\begin{document}
\title{Efficient Magic State Distillation by Zero-Level Distillation}

\author{Tomohiro Itogawa}
\affiliation{%
  Graduate School of Engineering Science, Osaka University, 1-3 Machikaneyama, Toyonaka, Osaka 560-8531, Japan
}
\author{Yugo Takada}
\affiliation{%
  Graduate School of Engineering Science, Osaka University, 1-3 Machikaneyama, Toyonaka, Osaka 560-8531, Japan
}
\author{Yutaka Hirano}
\affiliation{%
  Graduate School of Engineering Science, Osaka University, 1-3 Machikaneyama, Toyonaka, Osaka 560-8531, Japan
}%
\author{Keisuke Fujii}%
\affiliation{%
  Graduate School of Engineering Science, Osaka University, 1-3 Machikaneyama, Toyonaka, Osaka 560-8531, Japan
}%
\affiliation{%
  Center for Quantum Information and Quantum Biology,
  Osaka University, 1-2 Machikaneyama, Toyonaka 560-0043, Japan
}%
\affiliation{%
  RIKEN Center for Quantum Computing (RQC),
  Hirosawa 2-1, Wako, Saitama 351-0198, Japan
}%
\affiliation{%
  Fujitsu Quantum Computing Joint Research Division, Center for Quantum Information and Quantum Biology, 
  Osaka University, 1-2 Machikaneyama, Toyonaka, Osaka, 565-8531, Japan
}%

\date{\today}

\begin{abstract}
Magic state distillation (MSD) is an essential element for universal fault-tolerant quantum computing, which distills a high-fidelity magic state from noisy magic states using ideal (error-corrected) Cliﬀord operations.
For ideal Cliﬀord operations, it needs to be performed on the logical qubits and hence incurs a large spatiotemporal overhead, which is one of the major bottlenecks for the realization of fault-tolerant quantum computers (FTQCs).
Here we propose zero-level distillation, which prepares a high-fidelity logical magic state at the physical level, namely {\it zero level}, using physical qubits and nearest-neighbor two-qubit gates on a square lattice.
We develop a zero-level distillation circuit and show that distillation can be made even more eﬃcient than the conventional sophisticated approaches with logical level distillations.
The key idea involves the Knill {\it et al.}-type distillation using the Steane code and its careful mapping to the square-lattice architecture with error detection.
The distilled magic state on the Steane-code state is then teleported or converted to surface codes.
We numerically find that the error rate of the logical magic state scales as approximately $100 \times p^2$ in terms of the physical error rate $p$.
For example, with a physical error rate of $p = 10^{-4}$ ($10^{-3}$), the logical error rate is reduced to $p_L = 10^{-6}$ ($10^{-4}$), resulting in an improvement of 2 (1) orders of magnitude.
This contributes to reducing both space and time overhead for early FTQC as well as full-fledged FTQC combined with conventional multilevel distillation protocols.

\end{abstract}

\maketitle

\section{Introduction}
Quantum computers are expected to provide advantages in solving problems that are intractable for classical computers such as prime factorization~\cite{Shor}, linear system solver~\cite{harrow2009quantum}, and quantum chemistry~\cite{Alan}.
Significant experimental efforts have been dedicated to the realization of quantum computers based on various physical systems.
Among these, the superconducting system is one of the most promising candidates, and systems with 50-100 qubits have already been experimentally demonstrated~\cite{Google,IBM}.
These quantum computers are called noisy intermediate-scale quantum computers (NISQ)~\cite{Preskill2018quantumcomputingin}, whose noise level is still high and the number of qubits is still limited, are currently unable to run sophisticated quantum algorithms with theoretically proven quantum speedup.
Although NISQ-aware algorithms are being developed~\cite{VQA}, an ultimate solution to these problems is to protect quantum information through quantum error correction~\cite{shor1995scheme} to realize a fault-tolerant quantum computer (FTQC)~\cite{FTQC}.

Surface codes~\cite{KITAEV20032,bravyi} are one of the most promising approaches for a fault-tolerant quantum computer using superconducting qubits, since they can be implemented on a two-dimensional square lattice and have high noise resilience~\cite{Fowler,Raussendorf}.
In FTQC, the entire computation has to be performed fault tolerantly, with quantum information encoded in logical qubits.
While Clifford gates can be implemented relatively easily in a fault-tolerant manner, non-Clifford gates, such as the $T$ gate, are hard to execute fault tolerantly~\cite{Eastin-Knill}.
Therefore, magic state distillation (MSD)~\cite{bravyi2005universal} is employed to prepare a high-fidelity magic state $T\ket{+}$ from noisy ones, which is hence used to implement the $T$ gate via gate teleportation~\cite{gate-teleportation}.

While MSD is a crucial operation for achieving universal quantum computation, it requires a large number of qubits, which is an obstacle to the realization of FTQC \cite{Gidney2021howtofactor}.
This is because in most MSD protocols distillation is performed by using logical qubits with concatenating a QEC code, which features transversal implementations of the $H$ or $T$ gates.
In order to mitigate this, physical-level distillation protocols have been proposed with error detection~\cite{goto, chamberland2019fault, chamberland2020very}.
Physical-level distillation has great potential to reduce the physical overhead, and its effectiveness has been demonstrated in an experiment with trapped ions~\cite{postler2022demonstration}.
However, this physical-level approach has not been used for resource estimation because it employs all-to-all or complicated gate connectivity, and a concrete implementation on an architecture compatible with surface codes is lacking.

In this work, we propose {\it zero-level distillation} to prepare a high-fidelity logical magic state using physical qubits and nearest-neighbor two-qubit gates on a square lattice.
Here zero level refers to performing distillation entirely at the physical level instead of distillation at a higher level using logical qubits.
More precisely, in zero-level distillation, a magic state is initially encoded into the Steane code non-fault-tolerantly.
Then this noisy magic state is verified using the Hadamard test of the logical $H$ gate following the Knill {\it et al.}-type protocol~\cite{knill1998resilient}.
The logical magic state is teleported to the rotated or planar surface code.
In addition to the teleportation-based approach, we also develop a fault-tolerant code conversion from the Steane code to the surface code, which allows us to reduce the number of qubits further while the depth is increased.

Throughout these processes, the operations employed are single-qubit gates and measurements and nearest-neighbor two-qubit gates.
Furthermore, any single-point error is detected by syndrome measurements and the Hadamard test, and hence the logical error rate scales as $\mathcal{O}(p^2)$ with respect to the physical error rate $p$.
Since QEC codes are not needed for fault-tolerant logical Clifford gates, the spatial overhead is much smaller than that of  conventional MSD protocols.

It might be thought that the advantage of the proposed zero-level distillation is limited to be the case with very low physical error rate because of the square-lattice constraint.
However, this is not the case if the distillation circuit is carefully designed.
For performance analysis, we fully simulate zero-level distillation circuits consisting of approximately 40 qubits, reducing the number of qubits required for simulation to 23 qubits.
As a result, zero-level distillation reduces the logical error rate of a magic state $p_L$ to $p_L \simeq 100 \times p^2$ by using only a physical depth of 25 for the rotated surface code (23 for the planar surface code).
For example, in the case of $p=10^{-4}$ ($p=10^{-3}$), the logical error rate results in $p_L=10^{-6}$ ($p_L=10^{-4}$), while the success rate is fairly high, 70\% (95\%).
Although the $100p^2$ scaling is slightly worse than $35p^3$ for the conventional method, it offers practical advantages in terms of the overall spatiotemporal overhead.
Specifically, in the context of the early-FTQC architecture~\cite{akahoshi}, the zero-level distillation facilitates a significant enhancement, increasing the reliability of non-Clifford gate operations by 2 orders of magnitude, in contrast to a scenario without MSD.
For full-fledged FTQC, zero-level distillation combined with conventional multilevel distillation~\cite{Litinski2019magicstate} allows us to save the number of physical qubits significantly to achieve a given accuracy by reducing the number of levels.

The rest of the paper is organized as follows.
In Sec. \ref{preliminary}, we provide a preliminary explanation of the Steane code and MSD.
In Sec. \ref{zero level distillation}, we provide a detailed description of our proposal, zero-level distillation.
In Sec. \ref{conversion}, we explain the protocol to convert the Steane-code state to the rotated surface code directly.
In Sec. \ref{Numerical simulation}, we present the outcomes of our numerical simulation.
In Sec. \ref{discussion}, we discuss the implications of zero-level distillation to early-FTQC and full-fledged FTQC.
Then, Sec. \ref{conclusion} is devoted to a conclusion.

\section{PRELIMINARY}
\label{preliminary}
In this section, we provide a preliminary explanation of the Steane code and MSD based on the transversality of the Hadamard gate on it.

\begin{figure}[t]
  \centering
  \includegraphics[keepaspectratio, scale=1]{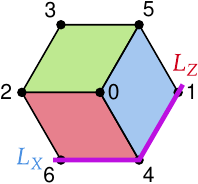}
  \caption{The Steane code. A qubit is located at each vertex.
  The red, blue, and green faces represent stabilizers.
  }
  \label{fig:steane_code}
\end{figure}

The Steane code is a $\llbracket 7, 1, 3\rrbracket$ stabilizer code,
which can correct an arbitrary single-qubit Pauli error
or detect an arbitrary two-qubit Pauli error.
Stabilizer generators are defined by $XIXIXIX, XXIIXXI, XIXXIXI,\\$$ZIZIZIZ, ZZIIZZI, ZIZZIZI$,
each corresponds to one of the three colored faces in
Fig.~\ref{fig:steane_code}.
On the Steane code, all Clifford gates can be implemented transversally, while non-Clifford gates such as the $T$ gate cannot.
Nonetheless, leveraging the transversality of the Hadamard $H$ gate, a special resource state, referred to as the magic state, can be prepared fault tolerantly as follows.

\begin{figure}[htbp]
  \centering
  \includegraphics[keepaspectratio, scale=0.55]{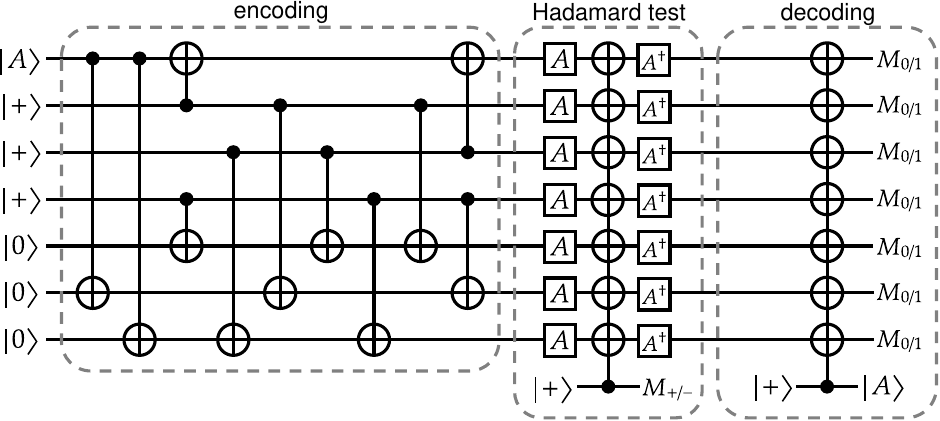}
  \label{fig:ALXA}
  \caption{MSD circuits based on transversality of the $H$ gate.
  The encoding circuit creates a logical magic state encoded with the Steane code.
The Hadamard test distills the magic state by measuring $H^{\otimes 7}$.
The decoding circuit is based on the one-bit teleportation.
}
  \label{fig:distillation}
\end{figure}

A magic state $\ket{A}$ is defined as $\ket{A} \equiv e^{-i\frac{\pi}{8}Y} \ket{+}$,
which is an eigenstate of $H$.
Leveraging the transversality of $H$ gate and controlled-Hadamard gates $\Lambda(H)$,
we implement a Hadamard test of the logical Hadamard operator
followed by decoding, as shown in
Fig.~\ref{fig:distillation}~\cite{knill1998resilient, knill1998resilient_2},
where the physical controlled-Hadamard gate $\Lambda(H)$ is
constructed by $A^\dagger \Lambda(X) A$ using the CNOT gate $\Lambda(X)$ and the non-Clifford gate $A$.
Suppose that all Clifford gates are ideal, and 
only non-Clifford gates are subject to errors with a probability $p$,
the above Hadamard test of the logical Hadamard operator and decoding process
detect up to two errors.
This provides the magic state with error rate $O(p^3)$.
In order to make Clifford gates ideal, each qubit in Fig.~\ref{fig:distillation} is further encoded into a logical qubit, and error-corrected Clifford gates are employed for distillation.
However, this approach costs a large amount of physical qubits and operations due to the concatenation of two QEC codes.

A lower-cost distillation method with flag qubits has been proposed by Goto \cite{goto}.
This protocol works with noisy Clifford gates by carefully designing the circuit with flag qubits 
so that any single-point error during the Clifford gates does not compromise the entire distillation process.
Since this approach eliminates the need for logical qubits in Clifford gates, 
it can substantially reduce the number of qubits needed for MSD.
However, the original proposal in Ref.~\cite{goto} relies on the Steane code and all-to-all gate connectivity,
which makes it difficult to apply for FTQC with the surface code
on a nearest-neighbor architecture.

\section{ZERO-LEVEL DISTILLATION}
\label{zero level distillation}
\begin{figure*}[htbp]
  \centering
  \includegraphics[scale = 0.4]{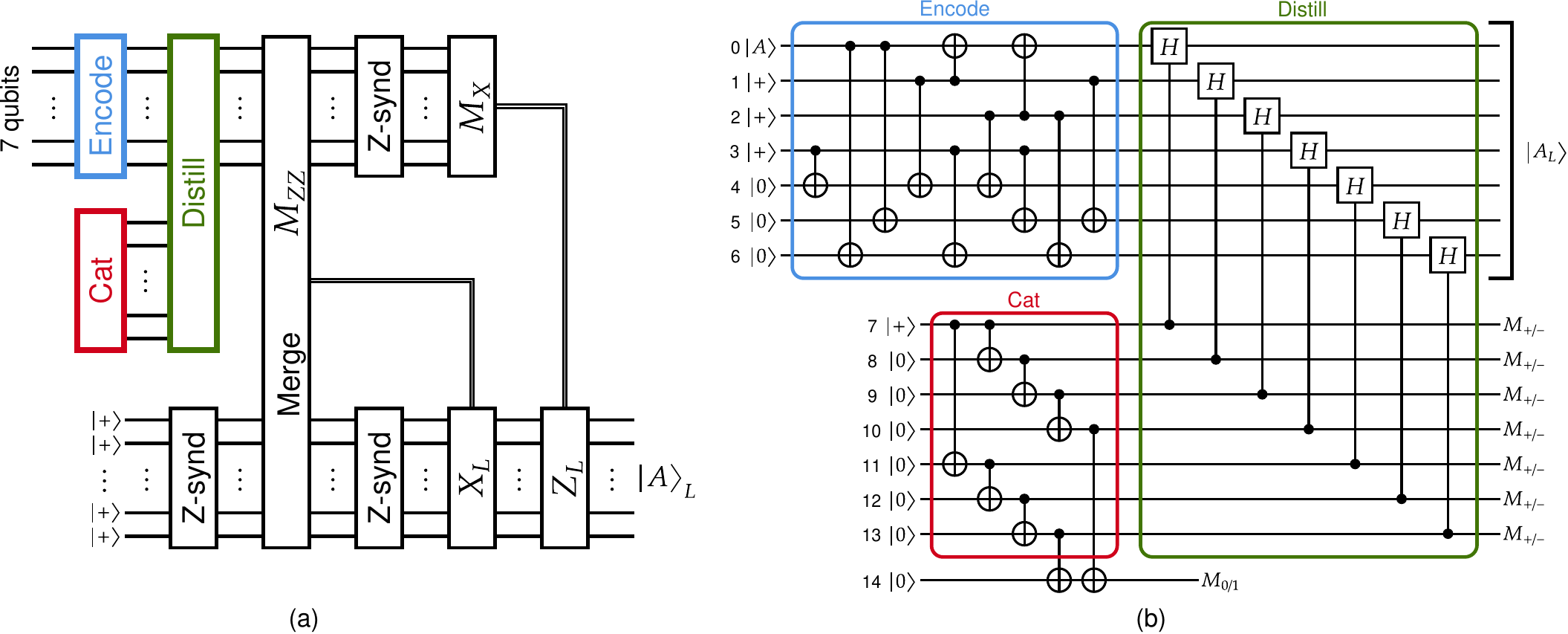}
  \caption{(a) The zero-level distillation circuit.
  (b) The detailed circuits for encoding of noisy magic state, preparation of cat state, and distillation.
  The circuit in the blue box encodes a magic state encoded with the Steane code non-fault-tolerantly.
  The circuit in the red box prepares a cat state.
  The circuit in the green box is the distillation circuit utilizing the Hadamard test.
  }
  \label{fig:zero_circuit}
\end{figure*}
In order to reduce the overhead of magic state distillation, we propose {\it zero-level distillation}.
This protocol prepares a logical magic state on the surface code without using multiple logical qubits.
We achieve physical-level distillation using the Steane code by carefully designing a fault-tolerant distillation circuit with fewer physical qubits and nearest-neighbor two-qubit gates on a square lattice.
Then, the logical magic state is teleported from the Steane code to the surface code,
which can be viewed as lattice surgery between color and surface codes~\cite{teleportation}.
(In the next section, we will also provide an alternative approach based on a code conversion from the Steane code to the surface code without teleportation.)
This combination of two QEC codes allows us to prepare 
the logical magic state using fewer physical qubits 
with high fidelity on a square lattice as detailed below.

Zero-level distillation consists of three key processes: non-fault-tolerant magic state encoding, postselection via the Hadamard test, and teleportation-based injection using lattice surgery.
The detailed steps  are as follows (see the schematic circuit diagram in Fig.~\ref{fig:zero_circuit}):

\begin{enumerate}[(i)]
  \item Encode a magic state in the Steane code non-fault-tolerantly. In parallel, a seven-qubit cat state is also prepared.
  \item Execute distillation by the Hadamard test using the cat state. 
  If the parity of the measurement outcome is even, the output state is accepted.
  In parallel, encode $\ket{+}_L$ with the rotated surface code.
  \item Merge and split the magic state and $\ket{+}_L$, and perform the projection by the logical $ZZ$ operator via the lattice surgery.
  \item Measure the $Z$ stabilizers on the Steane code and the surface code.
  \item Measure qubits on the Steane code directly in $X$ basis to complete teleportation.
\end{enumerate}

\begin{figure*}[htbp]
    \centering
    \includegraphics[keepaspectratio, scale=0.43]{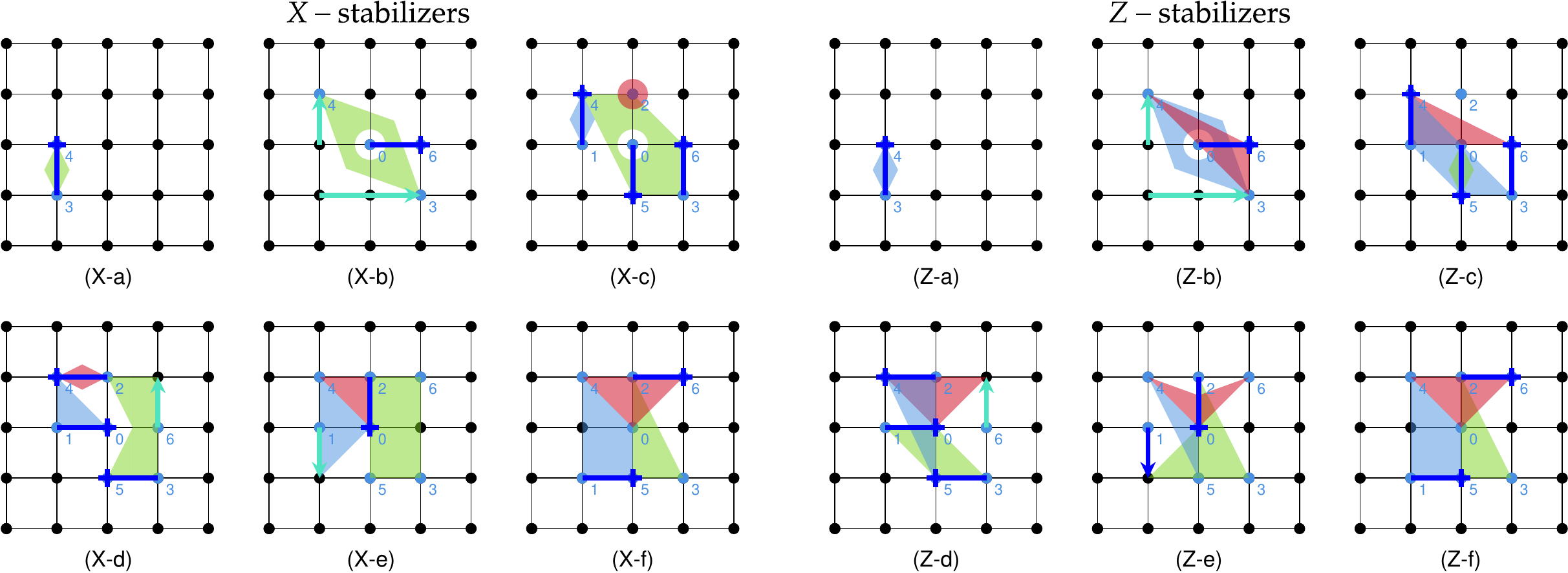}
    \caption{Qubit arrangement and stabilizer deformation during encoding a magic state.
    The left figure represents the $X$-stabilizer deformation, and the right figure illustrates the $Z$-stabilizer deformation.
    Each dot represents a qubit:
    the blue dots and numbers correspond to the locations of the qubits shown in Fig.~\ref{fig:zero_circuit}, and
    the light blue arrows indicate the transfer of qubits using one-bit teleportations.
    The blue bold edges indicate the CNOT gates, where the $+$ symbol on one side represents the target qubit.
    }
    \label{fig:steane_encode}
\end{figure*}

Fig.~\ref{fig:zero_circuit} shows the circuit for encoding a noisy logical magic state and preparing the cat state corresponding to step (i).
The blue box in Fig.~\ref{fig:zero_circuit} shows the circuit for non-fault-tolerant encoding of a magic state using the Steane code.
As shown in Fig.~\ref{fig:steane_encode}, the circuit can be constructed by using nearest-neighbor two-qubit gates on a square lattice.

\begin{figure*}[htbp]
    \centering
    \includegraphics[keepaspectratio, scale=0.55]{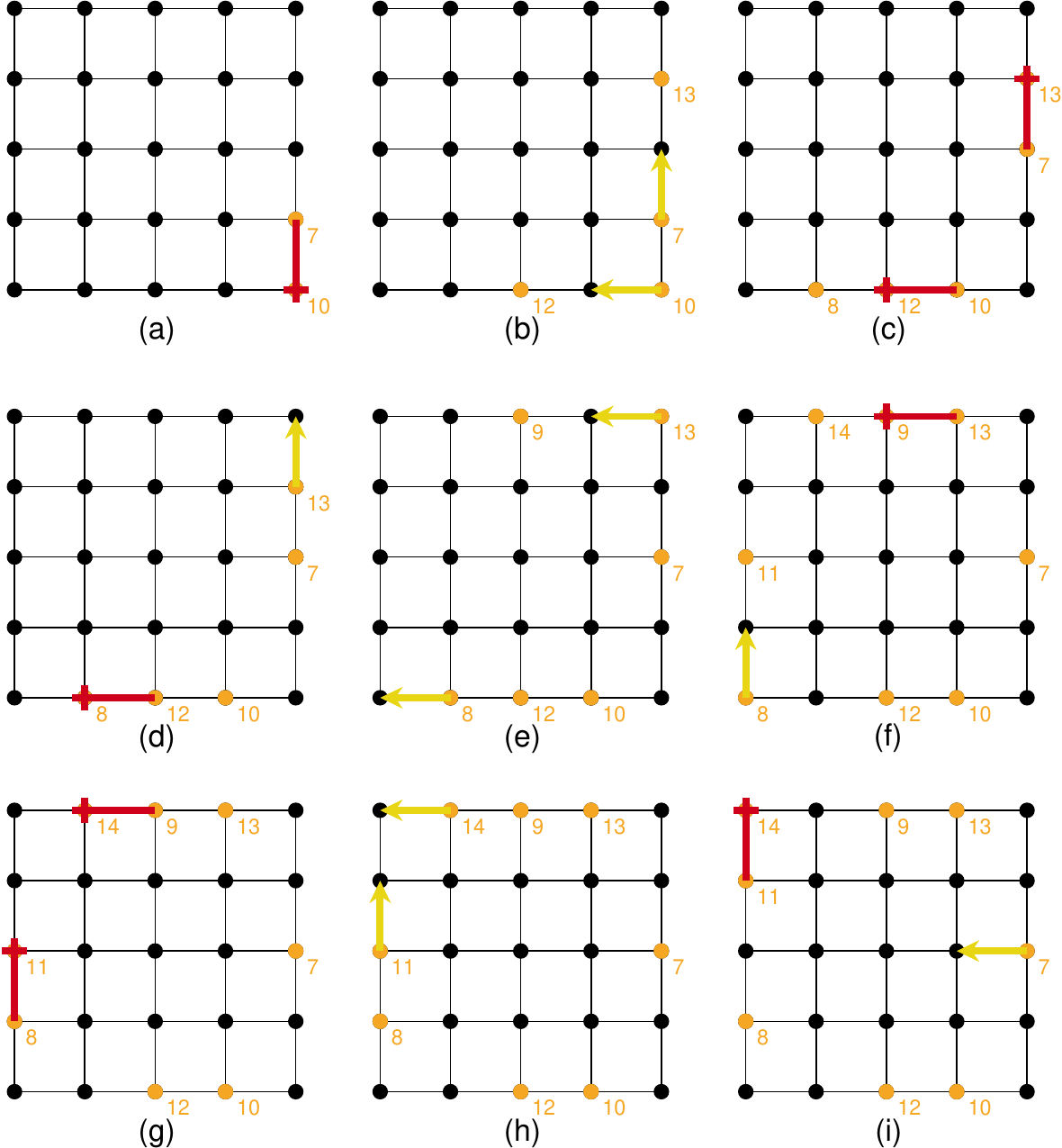}
  \caption{Qubit arrangement during encoding a cat state.
  Each dot represents a qubit:
  the orange dots and numbers correspond to the locations of the qubits in Fig.~\ref{fig:zero_circuit}.
  The yellow arrows indicate the transfer of qubits using one-bit teleportations.
  The red bold edges indicate the CNOT gates where the $+$ symbol on one side represents the target qubit.
  }
  \label{fig:cat_encode}
\end{figure*}

The red box in Fig.~\ref{fig:zero_circuit} shows the circuit for preparing a cat state.
While in previous work~\cite{goto}, magic state distillation is performed using two ancilla qubits, a seven-qubit cat state is more suitable in our situation where the qubit connectivity is highly limited.
Therefore, we use a seven-qubit cat state $\frac{1}{\sqrt{2}}(\ket{0}^{\otimes 7} + \ket{1}^{\otimes 7})$ as ancilla qubits.
This allows data and ancilla qubits to be adjacent on a square lattice as shown in Figs.~\ref{fig:steane_encode} and ~\ref{fig:cat_encode}.

Note that in Figs.~\ref{fig:steane_encode} and ~\ref{fig:cat_encode}, some qubits have to be moved so that CNOT gates between neighboring qubits can be performed.
Instead of using the swap operation, we employed one-bit teleportation to move a qubit to the neighboring site.
This is simply because the swap operation requires three CNOT gates, which results in a greater circuit depth.
However, if a swap gate can be implemented quickly and reliably, such as the $i$SWAP gate, it can be replaced with such a swap operation.

\begin{figure*}[t]
  \centering
  \includegraphics[scale = 0.6]{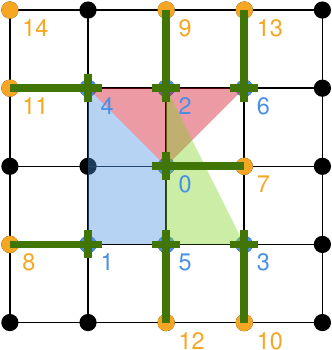}
  \caption{Qubit arrangement during the Hadamard test.
  The blue and orange dots and numbers indicate the locations of the qubits shown in Fig.~\ref{fig:zero_circuit}.
  A magic state and a cat state are encoded in the blue and orange qubits, respectively.
  The green edges indicate the CNOT gates, where the $+$ symbol on one side represents the target qubit.
  }
  \label{fig:distill}
\end{figure*}

\begin{figure*}[tbp]
  \centering
  \includegraphics[keepaspectratio, scale=0.43]{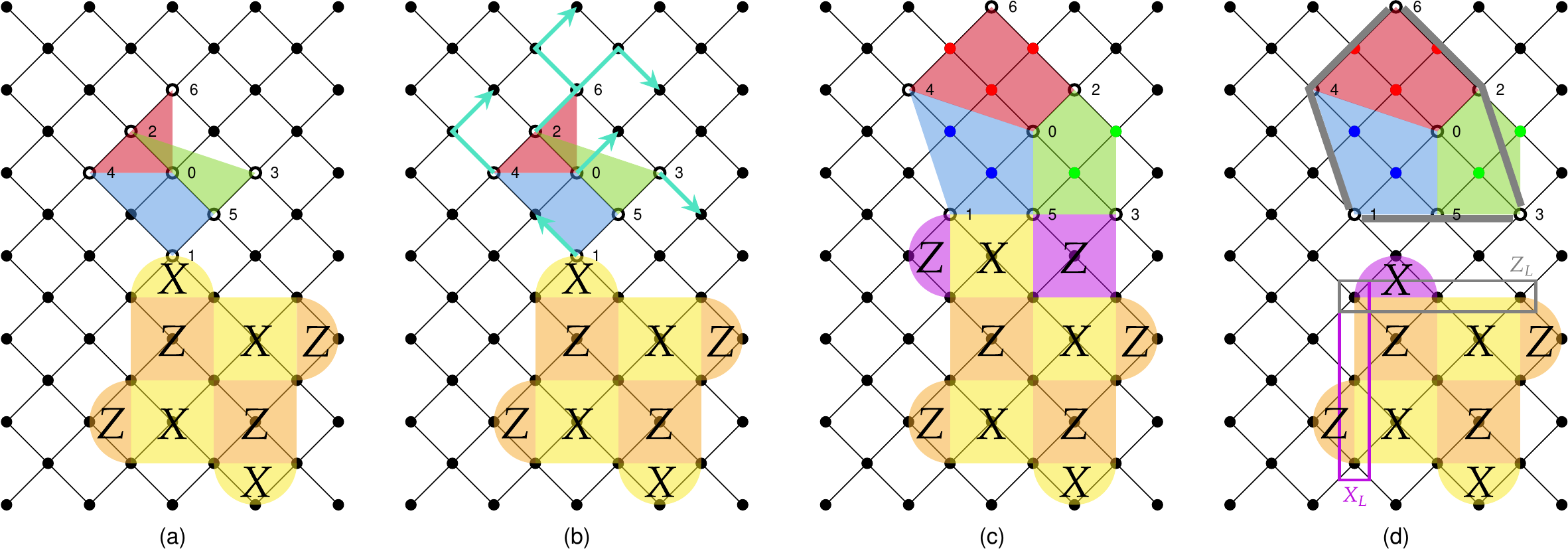}
  \caption{Qubit arrangement and measurements during teleportation from the Steane code to the rotated surface code.
  The white dots and numbers indicate the locations of the qubits shown in Fig.~\ref{fig:zero_circuit}, where
  the distilled magic state is encoded in the white qubits.
  The light blue arrows indicate the transfer of qubits using teleportation.
  The red, blue and green faces indicate the stabilizers of the Steane code, and the orange and yellow faces indicate the stabilizers of the rotated surface code.
  The purple faces indicate the $Z$ stabilizers for lattice surgery.
  The red, blue, and green dots are the ancilla qubits for the stabilizer measurements.
  The gray lines indicate logical operators of the Steane code.
  The purple box and the gray box indicate logical operators of the rotated surface code.
  }
  \label{fig:rotated_magic}
\end{figure*}

In step (ii), a Hadamard test for the logical Hadamard gate is performed using the cat state.
The green box in Fig.~\ref{fig:zero_circuit} shows the distillation circuit with the Hadamard test, where
the controlled-Hadamard gate $\Lambda(H)$ is implemented as $A^\dagger \Lambda(X) A$.
Fig.~\ref{fig:distill} (right) shows the details of the arrangement of qubits during distillation.
Since the data and ancilla qubits are positioned adjacent to each other, CNOT gates can be applied directly between them.
After the logical controlled-Hadamard gate,
the ancilla qubits from 7 to 13 in Fig.~\ref{fig:zero_circuit} are measured in the $X$ basis.
If the parity of the measurement outcomes is odd indicating any error, the distillation process is rejected.
Otherwise, the output is accepted and proceeds to the next step.
Next, some qubits are repositioned for the lattice surgery as shown in Fig.~\ref{fig:rotated_magic}(b).
Simultaneously, the $\ket{+}_L$ state is encoded using the rotated surface code by preparing $\ket{+}$ states and measuring the $Z$ stabilizers as shown in Fig.~\ref{fig:rotated_magic}(a).

In step (iii),
Steane and rotated surface codes are merged 
as shown in Fig.~\ref{fig:rotated_magic}(c), and then split via the lattice surgery.
This results in a projection onto $L_Z^{\mathrm{Steane}} \otimes L_Z^{\mathrm{Surface}}$ as its eigenvalue can be obtained from the product of $Z$-type operators at the boundary, indicated by the purple area of Fig.~\ref{fig:rotated_magic}(c).
During the lattice surgery, the $Z$ stabilizers must be measured twice to detect measurement errors.
If the measurement outcomes differ, the event is rejected, and the protocol must be restarted from the beginning.

\begin{figure}[htbp]
  \centering
  \includegraphics[keepaspectratio, scale=0.43]{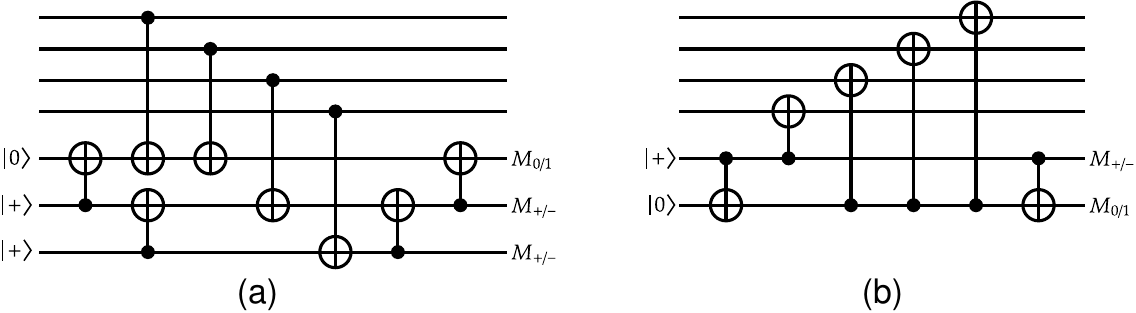}
  \caption{(a) The syndrome measurement circuit for the Steane code using three ancilla qubits. (b) The syndrome measurement circuit using two ancilla qubits. They correspond to red and blue-green in Fig.~\ref{fig:rotated_magic}(d). }
  \label{fig:rotated_synd}
\end{figure}

In step (iv), $Z$ stabilizers on the Steane code and the rotated surface code are measured.
Fig.~\ref{fig:rotated_synd} illustrates the syndrome measurement circuits for the Steane code~\cite{lao2020fault}, where the ancilla qubits in Fig.~\ref{fig:rotated_synd} correspond to red, blue, or green dots in Fig.~\ref{fig:rotated_magic}(d).
The measurements detect two-qubit errors introduced during merging, as well as errors during magic state encoding or $\ket{+}_L$-state encoding.

In step (v),
each qubit of the Steane code is directly measured in the $X$ basis to obtain the eigenvalues of the $X$ stabilizers and logical operators as shown in Fig.~\ref{fig:rotated_magic}(d).
Note that these eigenvalues should be interpreted appropriately according to the measurement outcome of the $X$ stabilizer at the boundary for splitting as usually done in the lattice surgery.
This completes the teleportation of the distilled magic state from the Steane code to the rotated surface code.
As usual, the Pauli frame of the logical qubit is updated based on the measurement outcome.
Throughout the entire procedure, a physical depth of 25 is used.
While we have demonstrated the proposed protocol using the rotated surface code,
 an implementation with the planar surface code is also provided in Appendix~\ref{planar}.


\begin{figure}[tbp]
  \centering
  \includegraphics[keepaspectratio, scale=0.36]{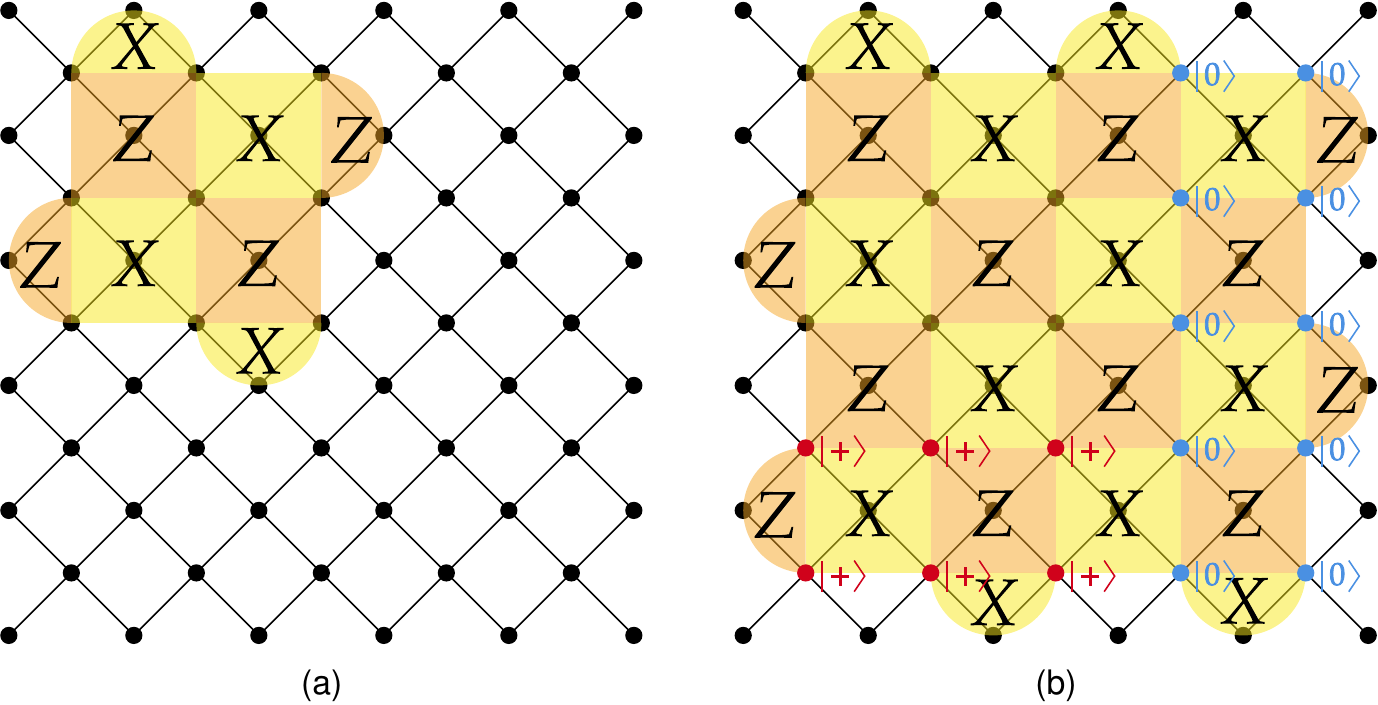}
  \caption{An expansion of the code distance from $d = 3$ to $d = 5$ in the rotated surface code. (a) Before expansion.
    (b) The output magic state on the $d=3$ surface code is located in the corner and other data qubits are initialized to $\ket{0}$ or $\ket{+}$ ancilla qubits.
    Then syndrome measurements are done, where error correction is performed without postselection.
  }
  \label{fig:extension}
\end{figure}

The above zero-level distillation generates a magic state encoded in a $d = 3$ surface code, but it is desirable that the output state is kept on the surface code with a larger code distance
to avoid error accumulation.
Therefore, the surface code is further expanded fault tolerantly as shown in Fig.~\ref{fig:extension}. 
In this process, $\ket{0}$ and $\ket{+}$ ancilla qubits are properly prepared~\cite{akahoshi} and stabilizer measurements are performed.

  \begin{figure*}[htbp]
  \centering
    \includegraphics[keepaspectratio, scale=0.43]{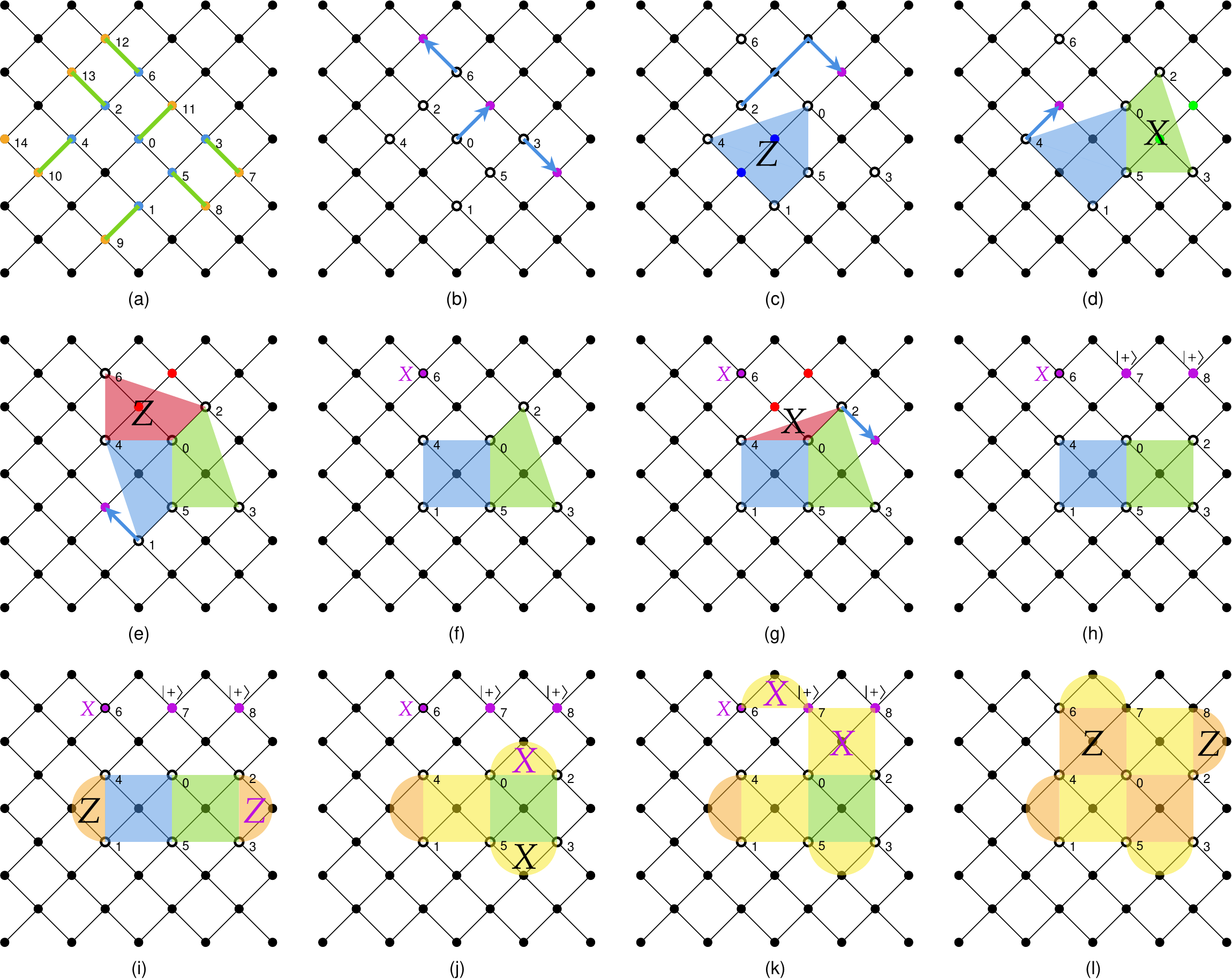}
    \caption{Qubit arrangement and measurements during conversion from the Steane code to the rotated surface code.
    The blue and orange dots and numbers correspond to the locations of the qubits shown in Fig.~\ref{fig:zero_circuit}.
    The blue arrows indicate the transfer of qubits using one-bit teleportations.
    The red, blue, and green faces indicate the stabilizers of the Steane code, and the orange and yellow faces indicate the stabilizers of the rotated surface code. The red, blue, and green dots are the ancilla qubits for the stabilizer measurements of the Steane code.
    }
    \label{fig:conversion}
  \end{figure*}
  
\section{DIRECT CODE CONVERSION FROM STEANE TO SURFACE CODES}
\label{conversion}
We propose an alternative protocol to prepare the logical magic state on the rotated surface code by converting the Steane-code state to the rotated surface code directly, without relying on teleportation.
This approach eliminates the need to generate two logical states and hence consumes fewer qubits compared to the teleportation-based approach described above.
As a drawback, the circuit depth is increased to 42, roughly double that of the teleportation method.
This leads to a higher logical error rate caused by additional idling noise.

First, as shown in Fig.~\ref{fig:conversion}(a), the magic state is encoded as described in Sec~\ref{zero level distillation}.
Subsequently, as shown in Fig.~\ref{fig:conversion} (b)--(l), we perform one-bit teleportations to move the qubits appropriately and 
stabilizer measurements to detect errors, and to form the stabilizer operators for the rotated surface code.
Specifically, Fig.~\ref{fig:conversion}(c)--(e) show $Z$-, $X$-, and $Z$-stabilizer measurements, respectively.
If any of the syndrome values is odd in these three stabilizer measurements, the distillation process must be restarted from the beginning.

Next, we deform the Steane code to a six-qubit code of distance two while preserving the logical state and fault tolerance.
As shown in Fig.~\ref{fig:conversion}(f),  qubit 6 is measured directly in the $X$ basis.
The six-qubit code consists of two weight-four $X$ and $Z$ stabilizers
and one weight-three $X$ stabilizer originating from the Steane code
as shown in Fig.~\ref{fig:conversion}(g).
Similar to the Steane code, the logical $Z$ operator is composed of three qubits, $Z_1 Z_3 Z_5$.
On the other hand, the logical $X$ operator is composed of two qubits, $X_1 X_4$.
In Fig.~\ref{fig:conversion}(h),  certain qubits are prepared in the $\ket{+}$ states,
which subsequently form part of an $X$-stabilizer in the rotated surface code.

Finally, the six-qubit code is deformed to the rotated surface code 
by measuring appropriate stabilizers with obtaining the syndrome values inherited from the Steane code.
In Fig.~\ref{fig:conversion}(i), two $Z$-type operators are measured; one is for the stabilizer of the rotated surface code and the other is to convert the $X$-type stabilizer of the Steane code (red face) into $X$ stabilizers of the rotated surface code.
The measurement outcomes of these two $Z$-type operators have to coincide since they are linked by the blue and green stabilizers of the Steane code.
Therefore, if the measurement outcomes disagree indicating any error, the distillation process is rejected.
Similarly, in Fig.~\ref{fig:conversion}(j),
two $X$-type operators are measured to form the $X$ stabilizers for the rotated surface code. If their measurement outcomes do not coincide, the distillation process is rejected. 
This measurement removes the $Z$ stabilizer on the blue face, and only the $X$ stabilizer is left (colored yellow).
In Fig.~\ref{fig:conversion}(k), the $X$ stabilizers are measured to be formed,
which is repeated twice to detect measurement errors.
In Fig.~\ref{fig:conversion}(l) two $Z$-type stabilizers are measured and formed,
which are repeated twice to detect measurement errors.
This measurement removes the $X$ stabilizer on the green face, and only the $Z$ stabilizer is left (colored orange).
This completes the code conversion, and the magic state is now encoded in the rotated surface code.
A physical depth of 42 is used to complete the code conversion, including the magic state distillation part on the Steane code.
Throughout this process, the Pauli frame should be updated based on the measurement outcomes as usual.

\section{NUMERICAL SIMULATION}
\label{Numerical simulation}
We perform numerical simulations to verify the fault tolerance of zero-level distillation and to estimate the resulting logical error rate and success rate.
Since our protocol includes non-Clifford gates,
we employ full-vector simulation using Qulacs~\cite{Suzuki2021qulacsfast}.
While the original circuits require approximately 40 qubits for the rotated surface code (50 for the planar surface code), we perform numerical simulations using only approximately 20 qubits 
by reusing the ancilla qubits without changing the structure of the original circuit~\cite{Katsuta}.
In the case of the code conversion, 15 qubits are enough for the entire simulation.

Regarding the noise model,
we employ a standard depolarizing noise model;
each single-qubit and two-qubit gate is followed by single-qubit and two-qubit depolarizing noise with a probability $p$, respectively,
where an idling process is also regarded as an identity gate and hence is followed by noise. 
Note that while this noise model has been widely used to evaluate the performance of existing fault-tolerant schemes, there is another noise model that is more realistic for superconducting qubit hardware such as SI1000~\cite{gidney2021fault}.
In such a noise model, it would be better to implement the qubit transportation with a direct swap operation to avoid measurements as mentioned before.
Also, we expect that applying SI1000 would lead to an improved logical error rate, since the zero-level distillation circuit includes quite a few idling operations, which have less impact with such an error model with much smaller idling noise.

In order to estimate the logical error rate $p_L$, 
the obtained magic state on the rotated surface code
is virtually projected to the code space, and fidelity between 
the obtained magic state and the ideal magic state is calculated. 
If the fidelity is not a unity, we count it as a logical error.
The reason why we project the state to the ideal code space 
is to estimate the logical error rate by excluding potentially detectable errors.
These detectable errors can be postselected in the further process in the leading order
by sacrificing a small amount of success rate.

\begin{figure}[tbp]
    \centering
    \includegraphics[scale=1]{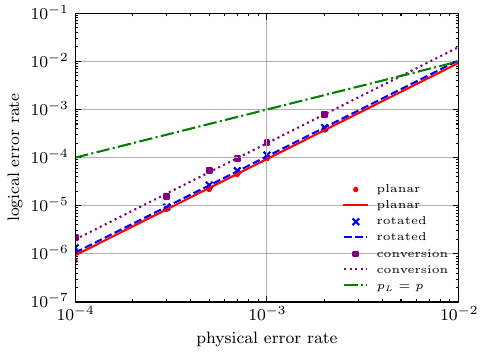}
    \label{fig:p2}
  \caption{The logical error rate $p_L$ plotted as a function of the physical error rate $p$ for the teleportation-based approaches of the planar (red circle) and rotated (blue cross) surface code, and the code conversion (purple square). The red, blue, and purple lines scale quadratically. The green line indicates $p_L = p$. }
  \label{fig:output}
\end{figure}

The logical error rate $p_L$ is estimated for various physical error rates
$p$,
where the numbers of samples is chosen to be between $10^6$ and $10^7$.
The resultant logical error rate is shown in Fig.~\ref{fig:output}
for the cases of teleportation-based approaches for the planar (red) and rotated (blue) surface codes, and the code-conversion approach (purple).
We can see that the logical error rate scales as $p_L = a \times p^2$ 
and the coefficient is $93.4$ for the planar surface code and 106 for the rotated surface code.
In the case of the code conversion, 
the coefficient is given by $199$.
Compared to the rotated surface code, the planar surface code 
has a slightly lower logical error rate due to less shuttling of qubits.
The intersection of $p_L \simeq 100 p^2$ and $p_L = p$ is located around $p = 10^{-2}$,
which indicates that the logical error rate is improved by 1 order of magnitude at $p = 10^{-3}$ and by 2 orders of magnitude at $p = 10^{-4}$.
Compared to the teleportation-based approach, 
the code conversion provides a slightly higher logical error rate because of its large circuit depth.

\begin{figure}[tbp]
  \centering
  \includegraphics[scale=1]{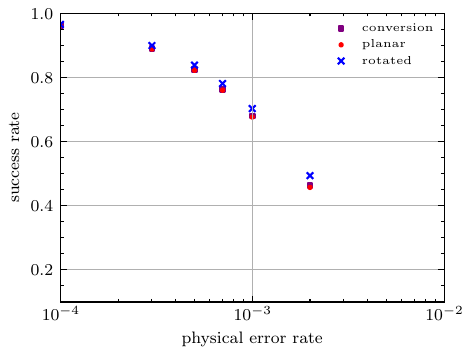}
  \caption{The success rate plotted as a function of the physical error rate $p$ for the teleportation-based approaches of the planar (red circle) and rotated (blue cross) surface code, and the code conversion (purple square).}
  \label{fig:success_rate}
\end{figure}

The success rate of distillation is plotted as a function of the physical error rate $p$ in Fig.~\ref{fig:success_rate}
for the teleportation-based approaches of the planar surface code (red) and the rotated surface code (blue), and the code conversion (purple).
The success rate is 70\% when $p = 10^{-3}$ and 95\% when $p = 10^{-4}$, indicating 
that distillation succeeds with a high probability.
Compared to the planar surface code, the rotated surface code has a slightly higher success rate since it uses fewer physical qubits.
The success rate of the code conversion is almost the same as that of the planar surface code since it uses fewer physical qubits and gates.


\begin{figure}[tbp]
    \centering
    \includegraphics[scale=1]{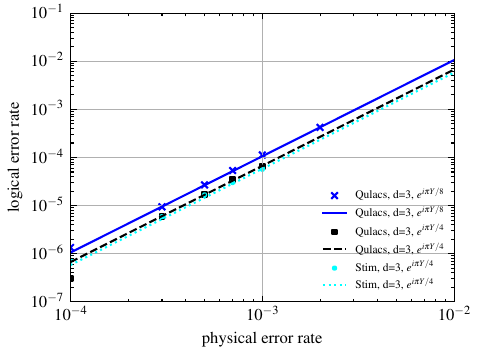}
    \caption{
    The logical error rate $p_L$ plotted as a function of the physical error rate $p$ for the $e^{-i \frac{\pi}{8} Y}$ distillation in Qulacs (blue cross), the $e^{-i \frac{\pi}{4} Y}$ distillation in Qulacs (black square), and the $e^{-i \frac{\pi}{4} Y}$ distillation in Stim (cyan dot).
    The blue, black, and cyan lines scale quadratically.
    }
    \label{fig:logical_physical_qs}
\end{figure}

\begin{figure}[tbp]
  \centering
  \includegraphics[scale=1]{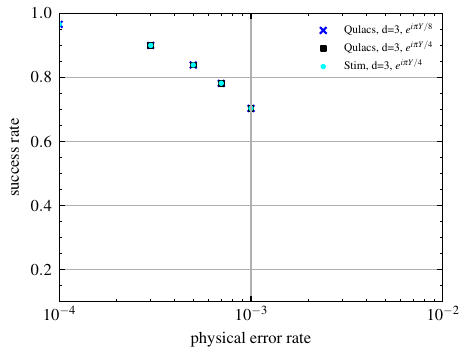}
  \caption{
    The success rate plotted as a function of the physical error rate $p$ for the $e^{-i \frac{\pi}{8} Y}$ distillation in Qulacs (blue cross), the $e^{-i \frac{\pi}{4} Y}$ distillation in Qulacs (black square), and the $e^{-i \frac{\pi}{4} Y}$ distillation in Stim (cyan dot).
  }
  \label{fig:success_rate_qs}
\end{figure}

We also perform numerical simulations to verify that surface-code expansion works fault tolerantly and to estimate the logical error rate and success rate in such a case.
For this purpose, a full-vector simulation using Qulacs cannot be used because of the large number of qubits.
Instead, we use the stabilizer simulator Stim~\cite{gidney2021stim}.
We replace all $e^{-i\frac{\pi}{8}Y}$ gates with $e^{-i\frac{\pi}{4}Y}$ gates and perform $e^{-i\frac{\pi}{4}Y}$ distillation instead of $e^{-i\frac{\pi}{8}Y}$ distillation.
Since this change does not affect the circuit configuration, it is reasonable to estimate the logical error rate.
In order to verify this,
we compare the results obtained by the full vector simulation
with Qulacs and Clifford simulation with Stim for the $d=3$ case without the expansion.
Fig.~\ref{fig:logical_physical_qs} shows the difference in the logical error rate between two simulation approaches for $e^{-i\frac{\pi}{4}Y}$ and $e^{-i\frac{\pi}{8}Y}$ distillations.
We find that replacing $\pi / 8$ with $\pi/4$ slightly decreases the logical error rate, but 
the success rate results in the same, as shown in Fig.~\ref{fig:success_rate_qs}.

\begin{figure}[tbp]
    \centering
    \includegraphics[scale=1]{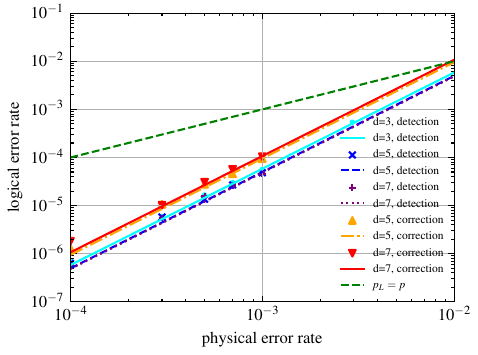}
  \caption{
  The logical error rate $p_L$ plotted as a function of the physical error rate $p$.
  Five data are plotted for each code distance and projection mode. The detection and correction modes mean performing error detection and correction, respectively when expanding the code.
Cyan, blue, purple, orange, and red lines scale quadratically.
The green line indicates $p_L = p$.
  }
  \label{fig:error_rate_stim}
\end{figure}

\begin{figure}[tbp]
  \centering
  \includegraphics[scale=1]{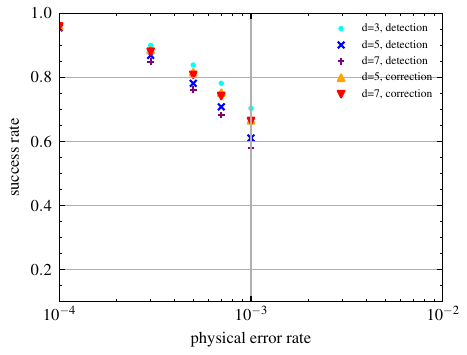}
  \caption{
  The success rate plotted as a function of the physical error rate $p$ for each code distance and projection mode. The detection and correction modes mean performing error detection and correction, respectively when expanding the code.
  }
  \label{fig:success_rate_stim}
\end{figure}

Fig.~\ref{fig:error_rate_stim} shows the results of the logical error rate when the code distance of an output state is extended.
The cyan dots indicate the logical error rate when the code distance is not extended, and other points (blue, purple, orange, red) indicate the data with the code distances extended from $d = 3$ to $d = 5$ or $d = 7$.
The cyan, blue, and purple points are calculated using error detection when projecting to the code space.
On the other hand, the orange and red points are calculated using error correction.
Fig.~\ref{fig:success_rate_stim} shows the results of the success rate when the code distance of an output state is extended.
As Fig.~\ref{fig:error_rate_stim}, five data points are plotted for each code distance and projection mode.
These results clearly demonstrate that the proposed method maintains a low logical error rate while slightly reducing the success rate, even when the surface code is expanded to mitigate the accumulation of logical errors after zero-level distillation.

\section{QUALITATIVE IMPLICATIONS OF ZERO-LEVEL DISTILLATION}
\label{discussion}
The potential implications of zero-level distillation include both early-FTQC~\cite{akahoshi} and full-fledged FTQC.
In early-FTQC, the limited availability of physical qubits prevents the execution of conventional multilevel distillation protocols~\cite{Litinski2019magicstate}.
Therefore, zero-level distillation, which operates with a very small number of qubits, spatial overhead of almost one logical qubit, is well suited for early-FTQC, despite its worse scaling $100p^2$ compared to $35p^3$ of conventional methods.
Zero-level distillation can generate magic states with a logical error rate of $10^{-6}$ under the physical error-rate assumption of $10^{-4}$, 
which would be sufficient to achieve Megaquop~\cite{preskill2025beyond} in early FTQC.
When applying continuous rotation gates with an approximation accuracy of $10^{-6}$, roughly speaking, a few tens of $T$ gates are required~\cite{ross2014optimal}.
By employing zero-level distillation, it becomes possible to perform around a few $10^4$ continuous rotation gate operations
with fully protected Clifford gates, thereby expanding the range of algorithms beyond NISQ.

Regarding the scalability to larger code distances for full-fledged FTQC applications, two studies, (0+1)-level distillation~\cite{hirano2024leveraging} and magic state cultivation~\cite{gidney2024magic}, have already been discussed
since the first version of this paper appeared.
(0+1)-level distillation utilizes the magic states generated by zero-level distillation as input states for conventional level-1 distillation protocols.
Combining zero-level distillation, whose logical error rate scales as $O(p^2)$, with conventional methods scaling as $O(p^3)$ results in an overall scaling of $O(p^6)$.
Zero-level distillation requires almost no additional qubits compared to conventional methods, resulting in a minimal increase in spatial overhead.
Notably, Ref.~\cite{hirano2024leveraging} has demonstrated an improvement in the logical error rate by up to 6 orders of magnitude using this method, while consuming the same spatiotemporal overhead $2.5\times 10^{5}$ qubitcycles.
Additionally, it is shown that the spatiotemporal overhead required to achieve a logical error rate of $10^{-16}$ can be reduced to approximately one third~\cite{hirano2024leveraging}.

On the other hand, magic state cultivation is inspired by the concept of zero-level distillation as mentioned in Ref.~\cite{gidney2024magic}. It employs larger color codes to perform physical-level distillation even under connectivity constraints. This method successfully suppresses the logical error rate to $O(p^5)$ by introducing various innovative techniques, such as the ``double-check" method. As a result, it can generate magic states with a given logical error rate while reducing the spacetime overhead by 2 orders of magnitude~\cite{gidney2024magic}.

\section{CONCLUSION}
\label{conclusion}
We proposed zero-level distillation, which efficiently distills and prepares the logical magic state encoded in surface codes without requiring multiple logical qubits.
All operations required can be implemented on the square lattice connectivity and the number of required physical qubits is substantially reduced and spatial overhead for one or two logical patches is sufficient.

According to the numerical simulation, zero-level distillation 
with teleportation successfully reduces the logical error rate $p_L$ of a logical magic state to $p_L=100 \times p^2$.
For example, when $p=10^{-3}$ and $p_L=10^{-4}$,
the logical error rates result in $p=10^{-4}$ and $p_L=10^{-6}$, respectively, indicating 1 and 2 orders of magnitude improvement.
The success rate is reasonably high even when $p=10^{-3}$.
The depth of the zero-level distillation circuit is only 25, hence it is compatible with the conventional multilevel distillation routines~\cite{Litinski2019magicstate}.
In addition, we also developed zero-level distillation based on the code conversion to further reduce the number of physical qubits employed.


\begin{acknowledgments}
  This work is supported by
  MEXT Quantum Leap Flagship Program (MEXT Q-LEAP) Grant JPMXS0120319794, JST COI-NEXT Grant No. JPMJPF2014, and JST Moonshot R\&D Grant No. JPMJMS2061.
\end{acknowledgments}

\appendix
\section{ZERO-LEVEL DISTILLATION FOR THE PLANAR SURFACE CODE}
\label{planar}
\begin{figure*}[tbp]
  \centering
  \includegraphics[keepaspectratio, scale=0.65]{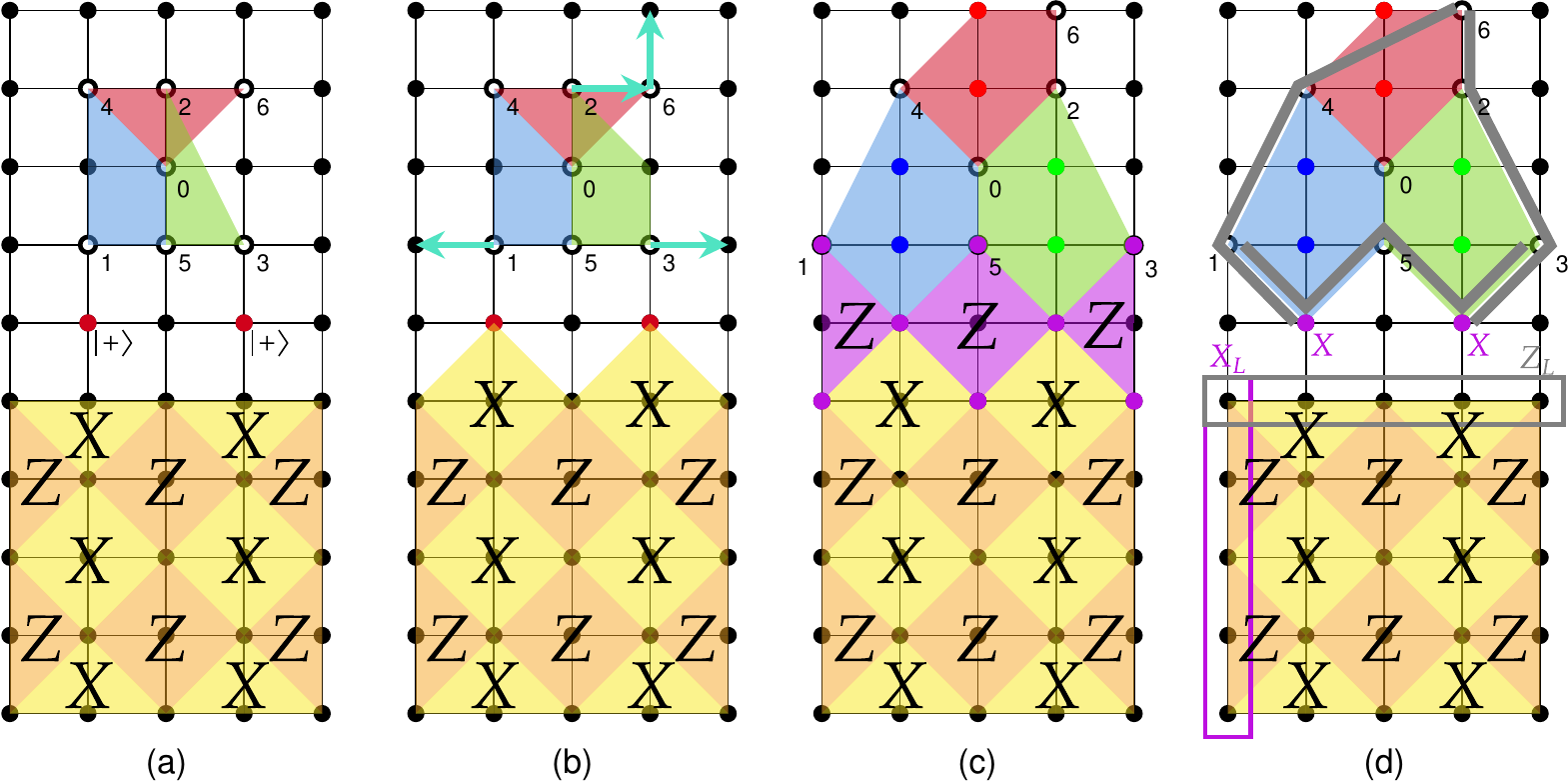}
  \caption{Qubits arrangement and measurements during teleportation from the Steane code to the planar surface code.
  The white dots and numbers correspond to the locations of the qubits shown in Fig.~\ref{fig:zero_circuit}, where
  the distilled magic state is encoded in the white qubits.
  The light blue arrows indicate the transfer of qubits using one-bit teleportations.
  The red, blue, and green faces indicate the stabilizers of the Steane code, and the orange and yellow faces indicate the stabilizers of the planar surface code.
  The red, blue, and green dots are the ancilla qubits for the stabilizer measurements of the Steane code.
  The purple faces indicate the $Z$ stabilizers at the boundary for the lattice surgery.
  The gray lines indicate logical operators of the Steane code.
  The purple box and the gray box indicate logical operators of the planar surface code.
  }
  \label{fig:planer_magic}
\end{figure*}

The magic state distilled on the Steane code can also be teleported to the planar surface code as well as the rotated surface code.
In this case, the $\ket{+}_L$ state encoded with the planar surface code and $\ket{+}$ ancilla physical qubits are prepared in parallel with the Hadamard test as shown in Fig.~\ref{fig:planer_magic}(a).
After the Hadamard test, some data qubits are moved as shown in Fig.~\ref{fig:planer_magic}(b).
Then, as shown by the purple area of Fig.~\ref{fig:planer_magic}(c), $L_Z^{\mathrm{Steane}} \otimes L_Z^{\mathrm{Surface}}$ is measured as a lattice surgery.
During this lattice surgery, the $Z$ stabilizers have to be measured twice to detect measurement errors.
As shown by the purple dots in Fig.~\ref{fig:planer_magic}(d), ancilla qubits on the boundary are measured in the $X$ basis, which breaks the $Z$ stabilizers on the boundary, and two code states are split.

As shown in Fig.~\ref{fig:planer_magic}(d), the $Z$ stabilizers are measured on both Steane and planar surface codes.
Finally, the qubits on the Steane code are directly measured in the $X$ basis to calculate the parities of $X$ stabilizers and logical operators.
Throughout all operations from the beginning, a physical depth of 23 is used.
Note that the Pauli frame has to be updated appropriately throughout these measurements as usual. 

\clearpage
\bibliographystyle{unsrt}
\bibliography{main}

\begin{thebibliography}{10}

\bibitem{Shor}
Peter~W. Shor.
\newblock Polynomial-time algorithms for prime factorization and discrete logarithms on a quantum computer.
\newblock {\em SIAM Journal on Computing}, 26(5):1484--1509, 1997.

\bibitem{harrow2009quantum}
Aram~W Harrow, Avinatan Hassidim, and Seth Lloyd.
\newblock Quantum algorithm for linear systems of equations.
\newblock {\em Physical review letters}, 103(15):150502, 2009.

\bibitem{Alan}
Alán Aspuru-Guzik, Anthony~D. Dutoi, Peter~J. Love, and Martin Head-Gordon.
\newblock Simulated quantum computation of molecular energies.
\newblock {\em Science}, 309(5741):1704--1707, 2005.

\bibitem{Google}
A.~Morvan et~al.
\newblock Phase transition in random circuit sampling, 2023.

\bibitem{IBM}
Youngseok Kim, Andrew Eddins, Sajant Anand, Ken~Xuan Wei, Ewout van~den Berg, Sami Rosenblatt, Hasan Nayfeh, Yantao Wu, Michael Zaletel, Kristan Temme, and Abhinav Kandala.
\newblock Evidence for the utility of quantum computing before fault tolerance.
\newblock {\em Nature}, 618(7965):500--505, Jun 2023.

\bibitem{Preskill2018quantumcomputingin}
John Preskill.
\newblock Quantum {C}omputing in the {NISQ} era and beyond.
\newblock {\em {Quantum}}, 2:79, August 2018.

\bibitem{VQA}
M.~Cerezo, Andrew Arrasmith, Ryan Babbush, Simon~C. Benjamin, Suguru Endo, Keisuke Fujii, Jarrod~R. McClean, Kosuke Mitarai, Xiao Yuan, Lukasz Cincio, and Patrick~J. Coles.
\newblock Variational quantum algorithms.
\newblock {\em Nature Reviews Physics}, 3(9):625--644, Sep 2021.

\bibitem{shor1995scheme}
Peter~W Shor.
\newblock Scheme for reducing decoherence in quantum computer memory.
\newblock {\em Physical review A}, 52(4):R2493, 1995.

\bibitem{FTQC}
Keisuke Fujii.
\newblock Quantum computation with topological codes: from qubit to topological fault-tolerance, 2015.

\bibitem{KITAEV20032}
A.Yu. Kitaev.
\newblock Fault-tolerant quantum computation by anyons.
\newblock {\em Annals of Physics}, 303(1):2--30, 2003.

\bibitem{bravyi}
Sergey Bravyi and Robert Raussendorf.
\newblock Measurement-based quantum computation with the toric code states.
\newblock {\em Phys. Rev. A}, 76:022304, Aug 2007.

\bibitem{Fowler}
Austin~G. Fowler, Matteo Mariantoni, John~M. Martinis, and Andrew~N. Cleland.
\newblock Surface codes: Towards practical large-scale quantum computation.
\newblock {\em Phys. Rev. A}, 86:032324, Sep 2012.

\bibitem{Raussendorf}
Robert Raussendorf and Hans~J. Briegel.
\newblock A one-way quantum computer.
\newblock {\em Phys. Rev. Lett.}, 86:5188--5191, May 2001.

\bibitem{Eastin-Knill}
Emanuel~Knill Bryan~Eastin.
\newblock Restrictions on transversal encoded quantum gate sets.
\newblock {\em Phys. Rev. Lett. 102, 110502}, 2009.

\bibitem{bravyi2005universal}
Sergey Bravyi and Alexei Kitaev.
\newblock Universal quantum computation with ideal clifford gates and noisy ancillas.
\newblock {\em Physical Review A}, 71(2):022316, 2005.

\bibitem{gate-teleportation}
Daniel Gottesman and Isaac~L. Chuang.
\newblock Demonstrating the viability of universal quantum computation using teleportation and single-qubit operations.
\newblock {\em Nature}, 402(6760):390--393, 1999.

\bibitem{Gidney2021howtofactor}
Craig Gidney and Martin Eker{\aa{}}.
\newblock How to factor 2048 bit {RSA} integers in 8 hours using 20 million noisy qubits.
\newblock {\em {Quantum}}, 5:433, April 2021.

\bibitem{goto}
Hayato Goto.
\newblock Minimizing resource overheads for fault-tolerant preparation of encoded states of the steane code.
\newblock {\em Scientific Reports volume 6, Article number: 19578}, 2016.

\bibitem{chamberland2019fault}
Christopher Chamberland and Andrew~W Cross.
\newblock Fault-tolerant magic state preparation with flag qubits.
\newblock {\em Quantum}, 3:143, 2019.

\bibitem{chamberland2020very}
Christopher Chamberland and Kyungjoo Noh.
\newblock Very low overhead fault-tolerant magic state preparation using redundant ancilla encoding and flag qubits.
\newblock {\em npj Quantum Information}, 6(1):91, 2020.

\bibitem{postler2022demonstration}
Lukas Postler, Sascha Heu$\beta$en, Ivan Pogorelov, Manuel Rispler, Thomas Feldker, Michael Meth, Christian~D Marciniak, Roman Stricker, Martin Ringbauer, Rainer Blatt, et~al.
\newblock Demonstration of fault-tolerant universal quantum gate operations.
\newblock {\em Nature}, 605(7911):675--680, 2022.

\bibitem{knill1998resilient}
Emanuel Knill, Raymond Laflamme, and Wojciech~H Zurek.
\newblock Resilient quantum computation.
\newblock {\em Science}, 279(5349):342--345, 1998.

\bibitem{akahoshi}
Yutaro Akahoshi, Kazunori Maruyama, Hirotaka Oshima, Shintaro Sato, and Keisuke Fujii.
\newblock Partially fault-tolerant quantum computing architecture with error-corrected clifford gates and space-time efficient analog rotations, 2023.

\bibitem{Litinski2019magicstate}
Daniel Litinski.
\newblock Magic {S}tate {D}istillation: {N}ot as {C}ostly as {Y}ou {T}hink.
\newblock {\em {Quantum}}, 3:205, December 2019.

\bibitem{knill1998resilient_2}
Emanuel Knill, Raymond Laflamme, and Wojciech~H Zurek.
\newblock Resilient quantum computation: error models and thresholds.
\newblock {\em Proceedings of the Royal Society of London. Series A: Mathematical, Physical and Engineering Sciences}, 454(1969):365--384, 1998.

\bibitem{teleportation}
Hendrik Poulsen~Nautrup, Nicolai Friis, and Hans~J. Briegel.
\newblock Fault-tolerant interface between quantum memories and quantum processors.
\newblock {\em Nature Communications}, 8(1):1321, 2017.

\bibitem{lao2020fault}
Lingling Lao and Carmen~G Almudever.
\newblock Fault-tolerant quantum error correction on near-term quantum processors using flag and bridge qubits.
\newblock {\em Physical Review A}, 101(3):032333, 2020.

\bibitem{Suzuki2021qulacsfast}
Yasunari Suzuki, Yoshiaki Kawase, Yuya Masumura, Yuria Hiraga, Masahiro Nakadai, Jiabao Chen, Ken~M. Nakanishi, Kosuke Mitarai, Ryosuke Imai, Shiro Tamiya, Takahiro Yamamoto, Tennin Yan, Toru Kawakubo, Yuya~O. Nakagawa, Yohei Ibe, Youyuan Zhang, Hirotsugu Yamashita, Hikaru Yoshimura, Akihiro Hayashi, and Keisuke Fujii.
\newblock Qulacs: a fast and versatile quantum circuit simulator for research purpose.
\newblock {\em {Quantum}}, 5:559, 2021.

\bibitem{Katsuta}
Mitsuki Katsuda, Kosuke Mitarai, and Keisuke Fujii.
\newblock Simulation and performance analysis of quantum error correction with a rotated surface code under a realistic noise model.
\newblock {\em Phys. Rev. Res.}, 6:013024, Jan 2024.

\bibitem{gidney2021fault}
Craig Gidney, Michael Newman, Austin Fowler, and Michael Broughton.
\newblock A fault-tolerant honeycomb memory.
\newblock {\em Quantum}, 5:605, 2021.

\bibitem{gidney2021stim}
Craig Gidney.
\newblock Stim: a fast stabilizer circuit simulator.
\newblock {\em {Quantum}}, 5:497, July 2021.

\bibitem{preskill2025beyond}
John Preskill.
\newblock Beyond nisq: The megaquop machine.
\newblock {\em ACM Transactions on Quantum Computing}, 6(3), April 2025.

\bibitem{ross2014optimal}
Neil~J. Ross and Peter Selinger.
\newblock Optimal ancilla-free clifford+t approximation of z-rotations.
\newblock {\em Quantum Info. Comput.}, 16(11–12):901–953, September 2016.

\bibitem{hirano2024leveraging}
Yutaka Hirano, Tomohiro Itogawa, and Keisuke Fujii.
\newblock Leveraging zero-level distillation to generate high-fidelity magic states.
\newblock In {\em 2024 IEEE International Conference on Quantum Computing and Engineering (QCE)}, volume~1, pages 843--853, Montréal, 2024. IEEE.

\bibitem{gidney2024magic}
Craig Gidney, Noah Shutty, and Cody Jones.
\newblock Magic state cultivation: growing t states as cheap as cnot gates.
\newblock {\em arXiv preprint arXiv:2409.17595}, 2024.

\end{thebibliography}

\end{document}